\documentclass[letterpaper]{article} 

\usepackage{booktabs}
\usepackage{dcolumn}
\usepackage{subcaption}
\usepackage{amssymb,amsmath}
\usepackage{cleveref}
\usepackage{xspace}
\usepackage{cleveref}

\usepackage{marginnote}

\newcommand\ca[0]{\emph{co-active}\xspace}

\newcommand\ro[0]{\emph{Reddit-only}\xspace}
\newcommand\subreddits[0]{r/The\_Donald, r/GenderCritical, and r/Incels\xspace}
\newcommand{\Act}{\text{Coactive}}

\usepackage{aaai23}  
\usepackage{times}  
\usepackage{helvet}  
\usepackage{courier}  
\usepackage[hyphens]{url}  
\usepackage{graphicx} 
\usepackage{natbib}  
\usepackage{caption} 
\usepackage{algorithm}
\usepackage{algorithmic}

\usepackage{newfloat}
\usepackage{listings}
\DeclareCaptionStyle{ruled}{labelfont=normalfont,labelsep=colon,strut=off} 
\floatstyle{ruled}
\newfloat{listing}{tb}{lst}{}
\floatname{listing}{Listing}
\title{Spillover of Antisocial Behavior from Fringe Platforms: The Unintended Consequences of Community Banning}
\author{
    Giuseppe Russo\textsuperscript{\rm 1}, Luca Verginer \textsuperscript{\rm 1}, Manoel Horta Ribeiro \textsuperscript{\rm 2}, Giona Casiraghi \textsuperscript{\rm 1}
}
\affiliations{
    \textsuperscript{\rm 1}ETH Zurich,
    \textsuperscript{\rm 2}EPFL\\
    giusepperusso@ethz.ch, lucaverginer@ethz,ch, manoel.hortaribeiro@epfl.ch, gionacasiraghi@ethz.ch

}

\usepackage{dcolumn,booktabs}
\newcolumntype{d}[1]{D{.}{.}{#1}}

\newcommand{\Period}{\text{Period}}
\newcommand{\engagement}[0]{engagement }
\usepackage[detect-all]{siunitx}

\usepackage{cleveref}
\usepackage{graphicx}
\usepackage[detect-all]{siunitx}
\usepackage{bibentry}

\usepackage{booktabs}
\usepackage{dcolumn}
\usepackage{subcaption}
\usepackage{amssymb,amsmath}
\usepackage{cleveref}
\usepackage{xspace}

\usepackage{dcolumn,booktabs}
\newcolumntype{d}[1]{D{.}{.}{#1}}
\newcommand\mc[1]{\multicolumn{1}{c}{#1}} 

\usepackage[detect-all]{siunitx}

\begin{document}

\maketitle
\begin{abstract}
The banning of problematic online communities from mainstream platforms like Reddit and Facebook is often met with enthusiasm as stakeholders 
 expect this will make platforms more civil and respectful. 
However, this policy can lead users to migrate to alternative fringe platforms with lower moderation standards and where antisocial behaviors like trolling and harassment are widely accepted.
As users of these communities often remain \ca across mainstream and fringe platforms, antisocial behaviors may spill over onto the mainstream platform.
We study this possible spillover by analyzing around  $70,000$ users from three banned communities that migrated to fringe platforms: r/The\_Donald, r/GenderCritical, and r/Incels.
Using a difference-in-differences design, we contrast \ca users with matched counterparts to estimate the causal effect of fringe platform participation on users' antisocial behavior on Reddit.
We find that participating in the fringe communities increases users' toxicity on Reddit (as measured by Perspective API) and involvement with subreddits similar to the banned community---which often also breach platform norms.
The effect intensifies with time and exposure to the fringe platform.
In short, we find evidence for a spillover of antisocial behavior from fringe platforms onto Reddit via co-participation.
\end{abstract}

\section{Introduction}
\label{sec:introduction}

\begin{figure}[t]
    \centering
    \includegraphics[width=0.825\columnwidth]{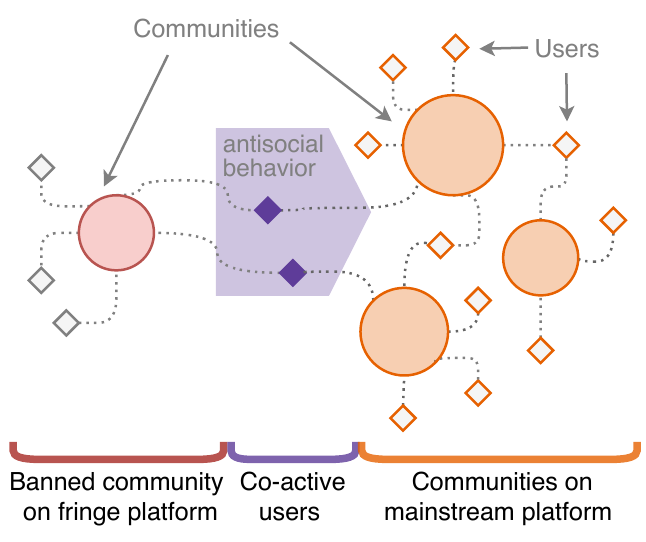}
    \caption{\textbf{Motivation.} When communities are banned from a mainstream platform and relocate to a fringe platform, antisocial behaviors may spill over onto the mainstream through \ca users, i.e., active across platforms.
    In this paper, we study this spillover effect by analyzing \ca users in three fringe communities banned from Reddit.}
    \label{fig:teaser}
\end{figure}

Online communities, ``aggregations of individuals who interact around a shared interest''~\cite{porter2004typology}, date back to the bulletin boards and chat systems of the early days of the Web~\cite{preece2003history}.
Today, thriving online communities are often hosted on mainstream social media platforms like Reddit and Facebook. 
Mainstream platforms moderate communities through a two-tiered governance system.
The platform is responsible for coarse-grained measures, like creating guidelines that all communities should adhere to and sanctioning communities that fail to conform to them~\cite{juneja2020through}.
On the community level, volunteer moderators make fine-grained moderation decisions, such as determining rules specific to the community and removing posts deemed inappropriate~\cite{seering2019moderator}.

Recently, online platforms have often banned---entirely deactivated---communities that breached their increasingly comprehensive guidelines. 
Importantly, such \emph{community-level bans} make a community unreachable without necessarily banning its individual users. Thus, members of a banned community may remain active on the platform.
In 2020 alone, Reddit banned around 2,000 subreddits (the name a community receives on the platform) associated with hate speech~\cite{reddit2020}.
Similarly, Facebook banned 1,500 pages and groups related to the QAnon conspiracy theory~\cite{qanon2020}. 
While these decisions are met with enthusiasm [e.g., see \citet{adl2020}], the efficacy of ``deplatforming'' these online communities has been questioned~\cite{zuckerman2021deplatforming}.
When mainstream platforms ban entire communities for their offensive rhetoric, users often migrate to alternative \emph{fringe platforms}, sometimes created exclusively to host the banned community~\cite{dewey_these_2015}.
Banning, in that context, would not only strengthen the infrastructure hosting these fringe platforms~\cite{zuckerman2021deplatforming} but allow these communities to become more toxic elsewhere~\cite{horta2021platform}.

Banning online communities may also impact the mainstream platforms themselves~\cite{trujillo2022make}. 
In \cref{fig:teaser}, we illustrate one mechanism by which this may happen.
When problematic communities are banned, users may choose to remain active in both mainstream and fringe platforms, creating feedback between online spaces with little to no moderation and social networks.
In the fringe platform, these \ca users are likely exposed to increased toxicity and misinformation and may participate in harassment, doxing, and defamation campaigns~\cite{freelon2020false}.
Consequently, antisocial behaviors from fringe platforms may spill over into other unbanned communities within mainstream social media where co-active users participate.

\paragraph{Present work}
In this paper, we conduct a large-scale longitudinal study of Reddit's users of banned communities.
We compare the post-ban behavior on Reddit of users that post exclusively on Reddit itself with that of users that post also on fringe platforms.
We find that users who co-participate---active on both platforms---exhibit more antisocial behavior on Reddit than users posting on Reddit only.
This effect intensifies over time and increases with activity (i.e., how much users write) on the fringe platform.
In short, we find spillovers of antisocial behavior from fringe platforms onto mainstream social media through \ca users.

\section{Related Work}
\label{sec:rel_work}

\paragraph{Measuring antisocial behavior on the Web.} 
Antisocial behavior has existed on the Web since its early days~\cite{dibbell1994rape}, with users engaging in different types of behavior like \textit{trolling,} i.e., intentionally disrupting a discussion or community~\cite{cheng2015antisocial}, and \textit{harassment}, attempts to demean or humiliate~\cite{PewHarassment}.
Previous works have attempted to measure the prevalence of antisocial behaviors~\cite{cheng2015antisocial, wulczyn2017ex}, as well as to understand factors that would lead users to engage in them~\cite{cheng2017anyone}.

One widely used machine learning tool to measure online antisocial behavior is Perspective API from Jigsaw~\cite{perspective}.
It provides  ``toxicity'' scores to posts indicating if they would lead to someone leaving a discussion due to their rude and disrespectful nature.
Perspective and other automated content moderation tools have faced widespread criticism: they lack context, fail to distinguish between legitimate and rule-breaking content, and are biased against minorities~\cite{vox,sap2019risk}.
At the same time, Perspective has proven to be a valuable tool for researchers to study online antisocial behavior.
Previous research on Reddit and Facebook data~\cite{rajadesingan2015sarcasm,kim2021distorting} shows that its performance is similar to that of a human annotator.
It further outperforms keyword-based alternatives~\cite{zannettou2020measuring}.

\paragraph{Online antisocial communities.} 

Antisocial communities are groups of users consistently engaging in antisocial behavior~\cite{marwick2018drinking}.
They are often sympathetic to conspiracy theories [e.g., QAnon~\cite{schulze2022far}] and extremist ideologies [e.g., the Alt-right~\cite{rieger2021assessing}].
They have been shown to have disproportionate influence over memes and news shared on the web~\cite{zannettou2018origins,zannettou2017web}.
Further, they have been closely associated with medical misinformation, conspiracy theories, and extremist ideologies that significantly impact the real world~\cite{zeng2021conceptualizing, mcilroy2019welcome,sipka2022comparing}.

Among these communities, the most relevant for this work are the following: r/The\_Donald, r/GenderCritical, and r/Incel.
The subreddit r/The\_Donald was created in June 2015 to support the then-presidential candidate Donald Trump's bid for the U.S. Presidential election.
This community has been closely linked with the rise of the ``alt-right'' movement, and was known to host racist, sexist and islamophobic discussions~\cite{lyons2017ctrl} and to spread conspiracy theories~\cite{paudel2021soros}.
\citet{flores2018mobilizing} have studied how active participants in r/The\_Donald mobilized the community to engage in ``political trolling''.
The subreddit r/GenderCritical was created in September 2013 to host the trans-exclusionary radical feminist (TERF) community.
TERFs hold the view that gender derives from biological sex~\cite{williams2020ontological}, and the community at large has consistently used social media to dox and harass trans women~\cite{atlantic}.
The subreddit  r/Incel was created in August 2013 to host a community of self-denominated ``\textit{in}voluntary \textit{cel}ibates.'' 
Incels abide by  ``The Black Pill,'' the belief that unattractive men would be doomed to romantic loneliness and unhappiness. 
Previous work has studied the community links with other masculinist communities~\cite{ribeiro2021evolution}, as well as its relationship with terrorist attacks~\cite{hoffman2020assessing} and the production of misogynistic content online~\cite{jaki2019online}.

\paragraph{Analyzing the effects of deplatforming.} Although different, a commonality between r/Incel, r/The\_Donald, and r/GenderCritical is that they have been ``de-platformed,'' i.e., banned from Reddit for breaching their guidelines.
Previous works have studied the effects of deplatforming of communities and users, finding that, following the ban, users reduce their activity on mainstream platforms~\cite{jhaver2021evaluating}, but also that users often migrate to other fringe platforms, where they at times become more toxic than before~\cite{horta2021platform,ali2021understanding}.
Moreover, \citet{trujillo2022make} have shown that users from banned communities may also become more toxic in other communities on the mainstream platform after the ban.

\paragraph{Relationship between prior and present work.} 
We analyze how co-participation in banned antisocial communities, now hosted in less moderated spaces, i.e., ``fringe'' platforms, increases antisocial behavior on the mainstream platform.
While previous work suggests that deplatforming may ``backfire'' due to creating more toxic communities on alternative platforms, we show that, additionally, antisocial behavior spills over onto mainstream platforms through co-active users.

\section{Data}

We use data from the three communities r/The\_Donald, r/Incels, and r/GenderCritical (see Section~\ref{sec:rel_work} for details).
In all three cases, after banning users migrated \textit{en masse} to alternative, fringe platforms (\emph{thedonald.win}, \emph{incels.co}, and \emph{ovarit.com}).
Thus, we collect the entire posting history consisting of both submissions and comments for the users active in these communities (i) on Reddit and (ii) on the relative fringe platform.

\paragraph{Reddit.} We collect all posts from Reddit through the Pushshift API \cite{baumgartner2020pushshift}. 
We collect all posts made on the three focal subreddits, starting eighteen weeks before they were banned. 
Specifically, for r/Incels, we collected data between July 20, 2017, and November 7, 2017; for r/The\_Donald, between November 11, 2019, and February 26, 2020; and for r/GenderCritical between February 14, 2020, and June 29, 2020. 
Overall, we collect four million posts from the three subreddits. 
Additionally, for each studied subreddit, we collect all contributing users' entire Reddit posting history.
To remove users with low activity in the banned subreddit [as commonly done in social computing research, see \citet{kumar2018mega} and \citet{samory2018conspiracies}], we consider only ``focal users,'' those with more than ten posts in the banned subreddit in the period prior to the banning.
Finally, to filter activity on small subreddits, we remove posts made in subreddits with less than five contributions from focal users.
The processed dataset contains  $181,787,627$ milion posts made on $72,991$ subreddits by  $69,970$ users ($61,569$ for r/The\_Donald, $5,367$ for r/GenderCritical, and $3,034$ for r/Incels).

\paragraph{Fringe Platforms.}
We implement and use custom web crawlers to collect data from \emph{thedonald.win}, \emph{incels.co}, and \emph{ovarit.com}, the fringe platforms where users of \emph{r/The\_Donald}, \emph{r/Incels}, and \emph{r/GenderCritical} respectively migrated following their ban.
For each platform, we collect all posts made eighteen weeks before and after the ban. 
We collect over $2.5$ million posts by $38,510$ users from \emph{thedonald.win},  $90,000$ posts by $1,560$ users from \emph{ovarit.com}, and $400,000$ posts by $2,270$ users from \emph{incels.co}.

\paragraph{Users labeling.}
To understand the effect of co-participation on fringe platforms on users' behavior on Reddit, we define \ca users as those posting both on Reddit and the fringe platforms after the banning. 
We track \ca users across platforms by exact string-matching their usernames.
Note that we assume that users with the same username across platforms correspond. 
A similar approach has been taken in previous work~\cite{horta2021platform,newell2016user}.
Note that \emph{r/The\_Donald} even had a system to facilitate username continuity across platforms~\cite{donaldpost1}.
Finally, we filter these users, keeping only those who made at least five posts on Reddit and the fringe platform after the ban and posted on the fringe platform \emph{only} after the ban.
We obtain $1,016$ Reddit users \ca on \emph{thedonald.win}, $176$ Reddit users \ca on \emph{ovarit.com}, and $286$ Reddit users \ca on \emph{incels.co}.

We label all users posting on Reddit without a matching username on the fringe platform as \ro users. 
We find  $10,829$ \ro users that were previously members r/The\_Donald, $1,228$ for r/GenderCritical \emph{Reddit-only}, and $2,753$ for r/Incels.
Finally, we match to each \ca user a \ro as described in \cref{sec:psm}.
This means that our final datasets contain 2032 users (1,016 \ca and 1,016 \ro) for r/The\_Donald, 352 users (176 \ca and  \ro) for r/GenderCritical, and 572 (286 \ca and \ro) for r/Incels.
We gather the activity of these users on a weekly basis, 18 weeks before and 18 weeks after the ban.
Our final datasets consist of $23158$, $1783$, and $3129$ observations for \subreddits, respectively.
A single observation in these datasets is given by the tuple \emph{(user, week, co-active, outcome variables)}.
Where \emph{week} is an integer in [-18, +18], \emph{co-active} is a variable indicating if a user is labeled as co-active as discussed above, and \emph{outcome variables} are described in \cref{subsec:outcome_variables}.

\section{Methods}\label{sec:methods}

To quantify the effect of co-participation on users' behavior on Reddit, we compare \emph{co-active} and \emph{Reddit-only} users.
We proxy antisocial behavior through users' toxicity (as measured through Perspective API) and their activity in other extreme subreddits (controversial group engagement).
To estimate the causal effects in observational data, we combine two widely used quasi-experimental causal inference methods: propensity score matching and difference-in-differences.

\begin{table}[t!]
\centering\small
\begin{tabular}{rp{5cm}}
\toprule
\multicolumn{2}{c}{\textbf{User Characteristics}}                                                                                                               \\ \midrule
\textbf{Participation}           & {Proportion of users' posts in the banned subreddit weighted by similarity.} \\
\textbf{Generality Score}    & {Activity diversity~\citep{waller2019generalists}}                       \\
\textbf{First Post Time}              & {Time of first post in the subreddit}                                            \\ \midrule
\multicolumn{2}{c}{\textbf{Language Characteristics}}                                                                                                           \\ \midrule
\textbf{Toxicity}                 & {A measure for usage of toxic language}                                                                   \\
\textbf{Anger and Anxiety}        & {Frequency of anger or anxiety words}                               \\ \midrule
\multicolumn{2}{c}{\textbf{Group Characteristics}}                                                                                                              \\ \midrule
\textbf{k-core centrality}                   & {Network embedness}           \\
\textbf{Eigencentrality}          & {Non-local network centrality}                                                         \\ \bottomrule
\end{tabular}
\caption{Description of the covariates used in the propensity score matching to ensure that \emph{Co-Active} and \emph{Reddit-Only} users are comparable. See \cref{app:propensity_score} for details.}\label{tab:predictors}
\end{table}

\subsection{Propensity Score Matching}\label{sec:psm}

We use a one-to-one propensity score matching to match \emph{co-active} and \emph{Reddit-only} users that were similar in the pre-banning period.
Propensity score matching (PSM) is a simple yet powerful method to account for selection bias that balances the distribution of observed covariates between groups.
This method allows us to mitigate the risk that observed differences in post-banning antisocial behavior exhibited by \emph{co-active} and \emph{Reddit-only} users come from user characteristics,
e.g., co-active users may be more toxic pre-banning and respond differently to the banning event.
PSM ensures that we consider users with equal probability to become active on the fringe platform.

PSM consists of three stages: (i) propensity score modeling, (ii) propensity score matching, and (iii) estimating a treatment effect after a successful balance check.
(i) We train a logistic regression classifier (LRC) to estimate the likelihood that a user will post on the fringe platform after the banning---the propensity score.
In particular, we trained the LRC on a set of user features computed on the pre-banning activities described in \cref{tab:predictors}. 
(ii)  We match each \emph{co-active} user to a \emph{Reddit-only} user using the nearest neighbor algorithm.
(iii) We test the quality of the matching by measuring the standardized mean difference of each covariate used in the PSM.
We obtained absolute standardized mean differences smaller than the standard $0.1$ threshold for all the covariates used to perform the PSM~\cite{austin2011introduction}.
We provide additional information about the robustness of the propensity score matching in \cref{app:propensity_score}.

\subsection{Difference-in-differences}\label{sec:did}
Considering the matched sample in the eighteen weeks before and after the ban date of each subreddit, we estimate the effect of co-activity in a fringe platform on users' behavior on  Reddit with the following difference-in-differences (DiD) model:
\begin{equation}
    \begin{split}\label{eq:base_did}
  Y_{it} =& \beta_0 + \beta_1 \Act_i + \beta_2 \Period_t + \\
       +&\beta_3 \Act_i \Period_t + u_i + \varepsilon_{it},
\end{split}
\end{equation}
where $Y_{it}$ is user $i$'s outcome (e.g., toxicity, we discuss outcomes in \cref{subsec:outcome_variables}) in period $t$ on Reddit.
$\Act_i$ indicates if user $i$ is \emph{co-active} or not.
$\Period_t$ indicates if the current time $t$ is before or after the ban ($t=0$).
$u_i$ is a fixed effect for user $i$ and $\varepsilon$ is the error term.
Under the assumption that the difference in outcomes between \ca and \ro users is constant over time in the absence of co-participation on fringe platforms (the ``parallel trends assumption'' the coefficient $\beta_3$ captures the causal effect of co-participation in the fringe community on the outcome variable.
Further, we ensure that $\beta_3$ does not incorporate self-selection effects.
In other words, we ensure that user-level characteristics such as demographics or political beliefs do not influence the likelihood of increased toxicity on the mainstream platform. 
Given the short duration of the observation window, we can assume such characteristics to remain constant in our time frame (36 weeks). 
Thus, by adding users' fixed effects $u_i$, we control for any biases introduced by user-level characteristics (e.g., demographics, political beliefs), mitigating the risk of omitted variable bias~\cite{bruderl2015fixed}. 

\subsection{Outcome Variables}\label{subsec:outcome_variables}

\paragraph{Toxicity.}
Previous works have shown how subreddits like \subreddits are prone to toxic language use~\cite{horta2021platform}.
These subreddits are home to many antisocial behaviors, such as incivility, harassment, trolling, and cyberbullying.
In this direction, the work of \citet{grover2019detecting} is particularly relevant, as it suggests that antisocial behaviors may be captured through automated text analysis.
Therefore, we employ the Perspective API~\cite{perspective} to measure the toxicity level of users' posts; see \cref{sec:rel_work} for details.
To infer a user's $i$ toxicity, we compute the median toxicity score $T_{it}$ of all the user's posts within a given time window $t$.
Specifically, for each user, we group their posts in weekly time windows to obtain weekly toxicity scores.

\paragraph{Engagement in controversial subreddits.}

We measure users' engagement with other controversial communities on Reddit as a second proxy for antisocial behavior.
For each user $i$, we compute the number of posts made in subreddits hosting discussions similar to the banned subreddit during a time window $t$.
We normalize this number by the total number of posts on the whole Reddit made by $i$ in the same time window.
We refer to the resulting measure $E_{it}$ as \engagement:
\begin{equation}
    E_{it} = \frac{\sum_{s \in S_K}{\lvert \lvert  P_{it}^{s}\rvert \rvert}}{\sum_{s \in R}{\lvert \lvert  P_{it}^{s}\rvert \rvert}}\,,
    \label{eq:affinity}
\end{equation}
where $S_K$ is the set of the k-th most similar subreddits to either \subreddits, $R$ is the set of all subreddits in Reddit (excluding the focal ones), 
and $\lvert \lvert P_{it}^{s} \rvert \rvert$ is the number of posts made by user $i$ at time $t$ in subreddit $s$.

To find the $k$-th most similar subreddits to \subreddits, we create a similarity scale in the interval [-1, +1] where 1 represents the highest similarity to the focal subreddit.
After manual annotation, we set k=$50$.
We provide further details about the similarity scale and manual annotation in \cref{sec:appendix}.

\section{Results}\label{sec:results}

We combine a large-scale longitudinal and regression analyses to assess the effect of co-participation in antisocial fringe platforms on users' behavior on Reddit.
We find that users co-participating on Reddit and fringe platforms exhibit increased antisocial behavior following a community ban.
More importantly, our study finds that the antisocial behavior of \ca users \emph{diverges over time} from that of \ro users.
We perform this analysis with two measures of antisocial behavior: (i) language toxicity and (ii) engagement with other controversial subreddits.
Our results are consistent for both measures across all the studied communities.

\begin{figure*}[t!]
    \centering
    \begin{subfigure}{.33\linewidth}\centering
    \caption{r/The\_Donald}\label{fig:data:thedonald}
    \includegraphics[scale=.6]{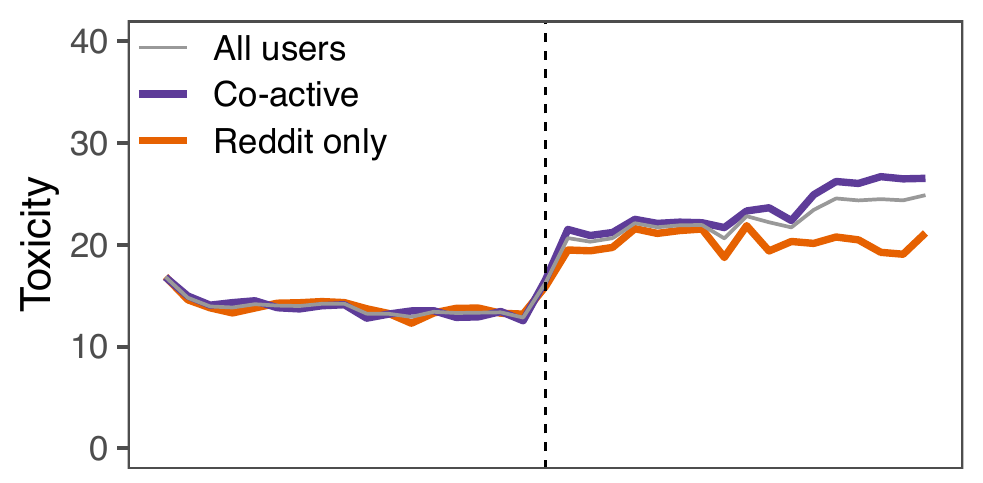}\\
    \includegraphics[scale=.6]{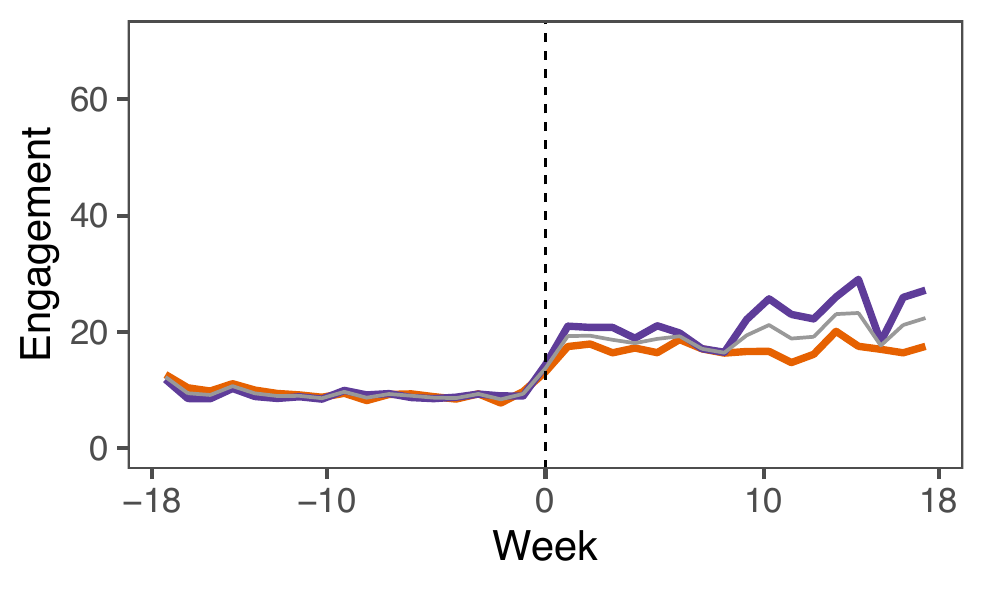}
    \end{subfigure}
   	\begin{subfigure}{.33\linewidth}\centering
   	\caption{r/GenderCritical}\label{fig:data:gc}
    \includegraphics[scale=.6]{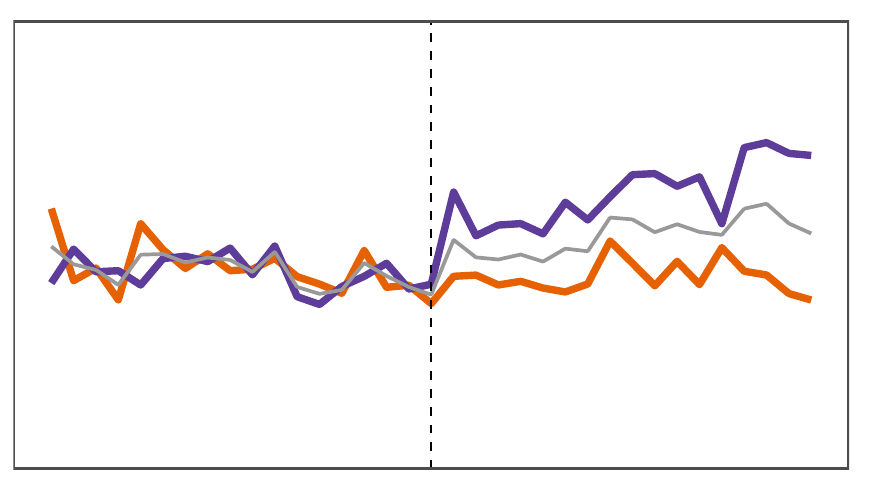}\\
    \includegraphics[scale=.6]{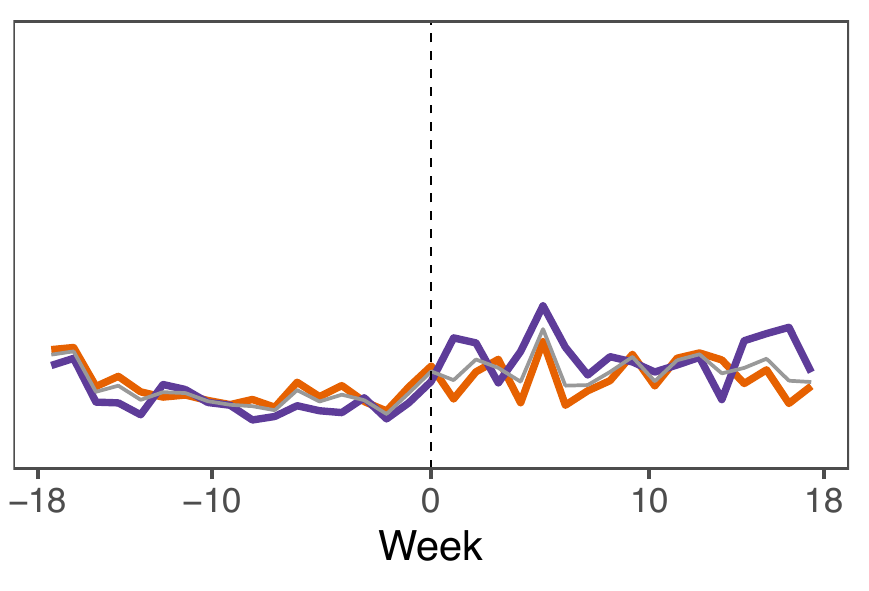}
    \end{subfigure}
	\begin{subfigure}{.33\linewidth}\centering
	\caption{r/Incels}\label{fig:data:incels}
    \includegraphics[scale=.6]{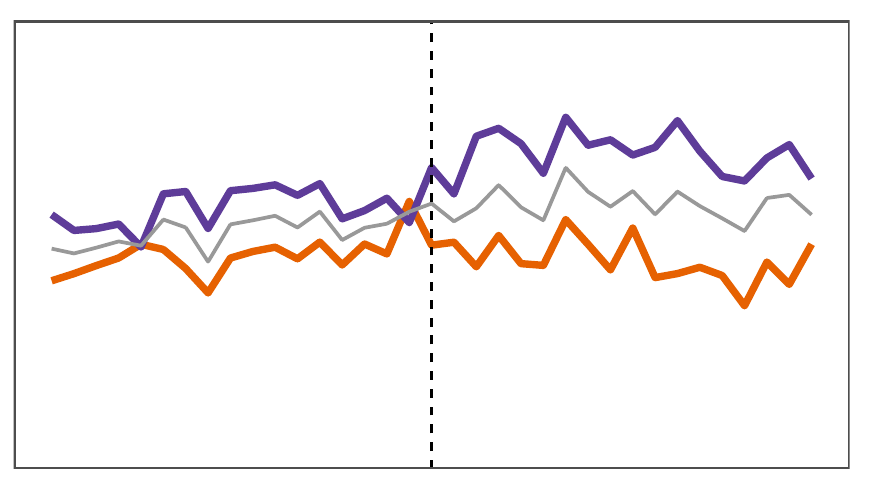}\\
    \includegraphics[scale=.6]{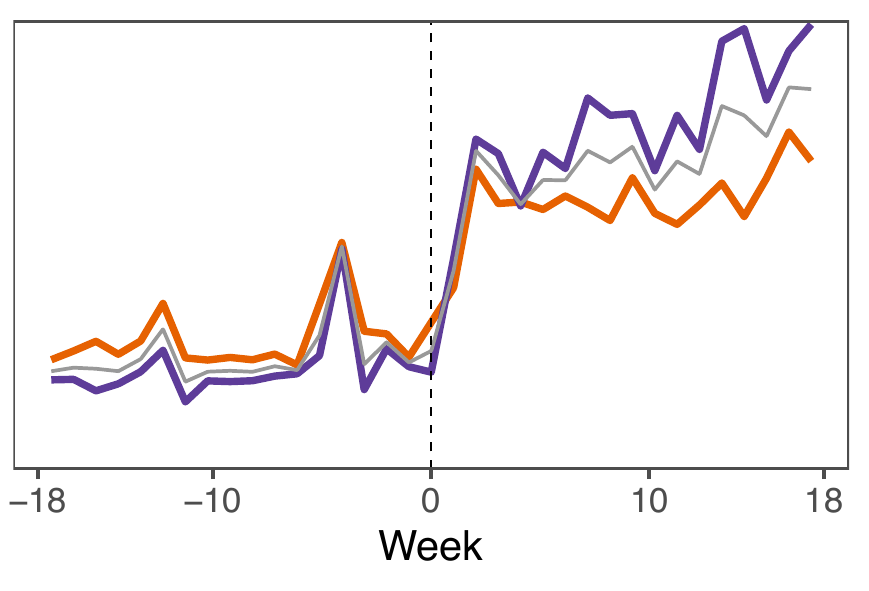}
    \end{subfigure}
    \caption{Toxicity mean values (top row) and Engagement mean values (bottom row) for \ca,  \ro, and \emph{all} users (purple, orange and grey lines, respectively).
     Toxicity and Engagement were computed over 36 weeks around the ban at Week$=0$, for r/The\_Donald (\cref{fig:data:thedonald}), r/GenderCritical (\cref{fig:data:gc}), and r/Incels (\cref{fig:data:incels}). }
    \label{fig:data}
\end{figure*}

\subsection{Longitudinal Analysis}

The upper row of \cref{fig:data} shows the toxicity of posts written on Reddit by users of r/The\_Donald, r/GenderCritical, r/Incels\xspace before and after the ban.
Note that here we consider the matched sample obtained after propensity score matching.
Following the ban, we observe that all users increase their toxicity (\cref{fig:data} grey line). 
The post-banning average toxicity grows by up to 61\% of the pre-banning toxicity (from 13 to 22).
This observation confirms the finding by \citet{trujillo2022make} of a marked increase in toxicity in the aftermath of community bans. 
By comparing \ca and \ro users separately (purple and orange lines, respectively), we see that the toxicity of \ca users grows faster than that of \ro.
For r/The\_Donald, \cref{fig:data:thedonald} shows a 68\% average increase in \ca users' toxicity after the ban.
This is a net increment of 23\% compared to \ro users. 
Similarly, we find a net increment of 41\% and 20\% for r/GenderCritical and r/Incels, respectively.

Qualitatively similar conclusions can be drawn when observing the \engagement of \ca and \ro users.
For instance, in the bottom row of \cref{fig:data}, we observe that \ca users of r/The\_Donald and r/Incels exhibit a steady increase in engagement towards controversial communities.
In particular, \ca users of r/The\_Donald increase their engagement from 9 to 19, while in r/Incels, they go from 15 to 50.
The case of r/Incels is particularly interesting, as the community remained active both on Reddit and in the subreddit r/braincels.
This subreddit gave continuity to members of r/Incels as they maintained their antisocial behavior habits. 
However, even under these circumstances \ca users show  more engagement towards Incels-related content than \ro users (see \cref{fig:data:incels} bottom)

\subsection{Difference-in-difference Analysis}

0\begin{figure*}[t!]
    \centering
    \begin{subfigure}{.33\linewidth}\centering
    \caption{r/The\_Donald}\label{fig:did:thedonald}
    \includegraphics[scale=.6]{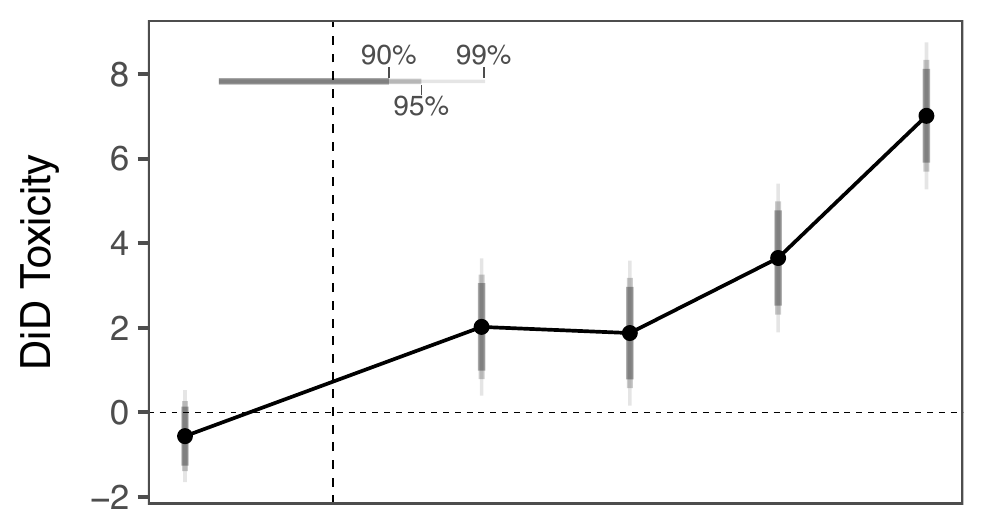}\\
    \includegraphics[scale=.6]{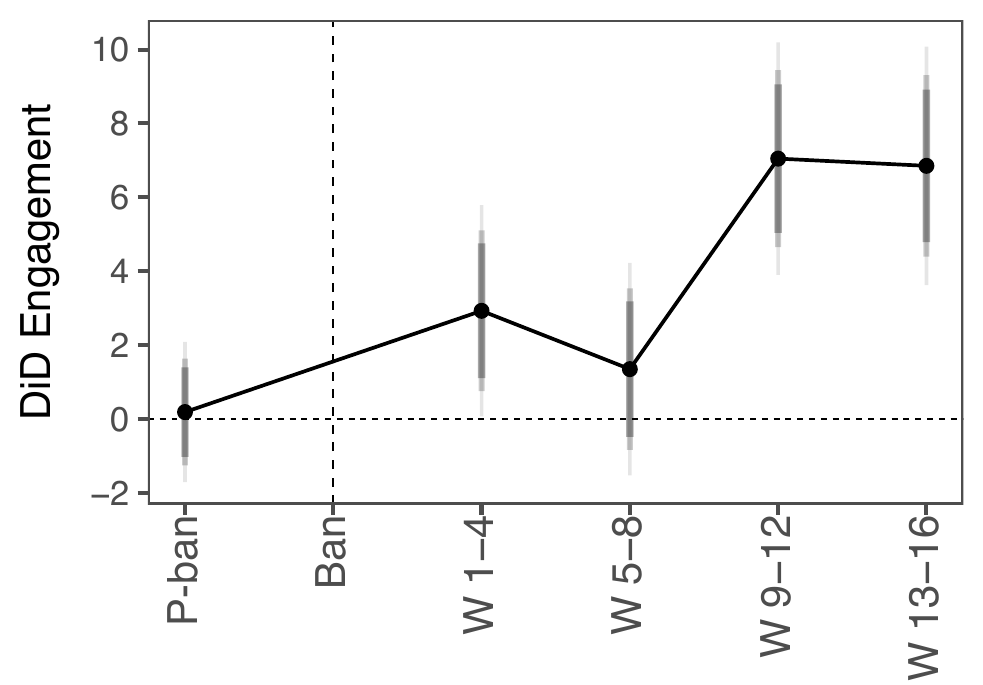}
    \end{subfigure}\hfill
   	\begin{subfigure}{.33\linewidth}\centering
   	\caption{r/GenderCritical}\label{fig:did:gc}
    \includegraphics[scale=.6]{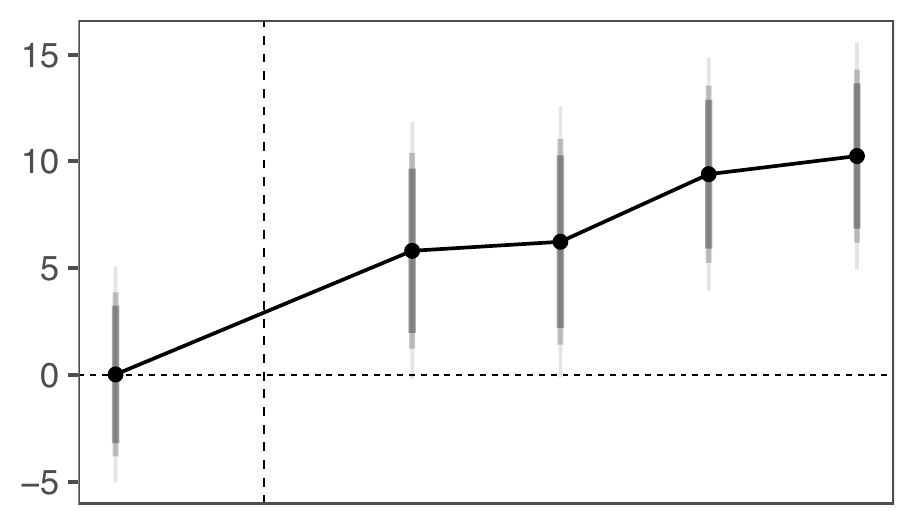}\\
    \includegraphics[scale=.6]{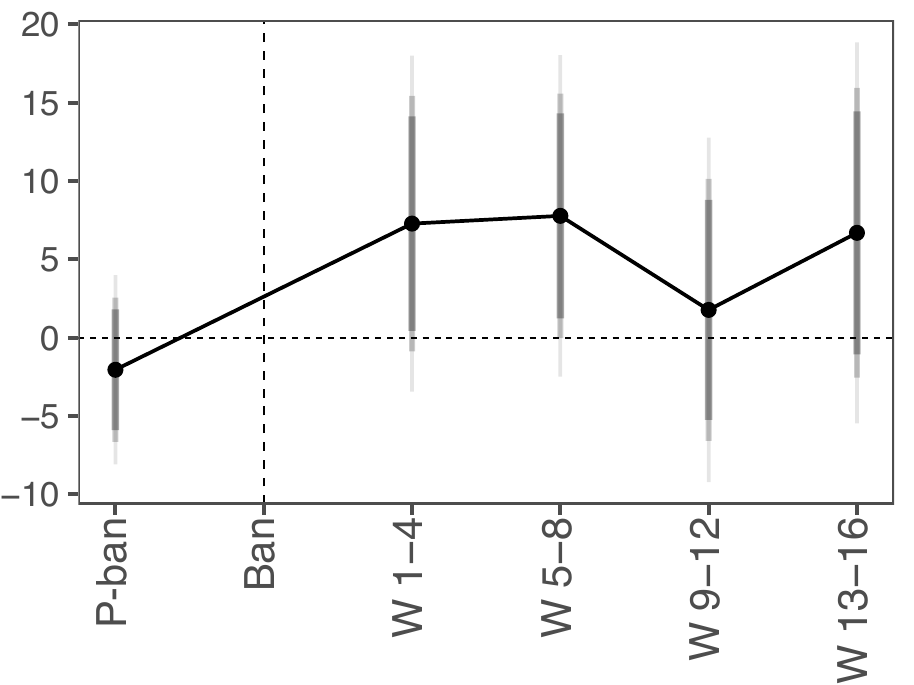}
    \end{subfigure}\hfill
	\begin{subfigure}{.33\linewidth}\centering
	\caption{r/Incels}\label{fig:did:incels}
    \includegraphics[scale=.6]{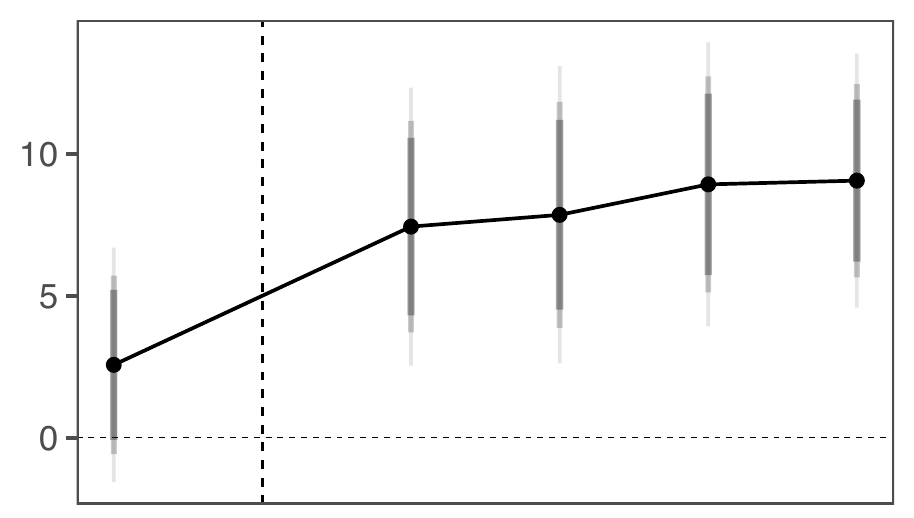}\\
    \includegraphics[scale=.6]{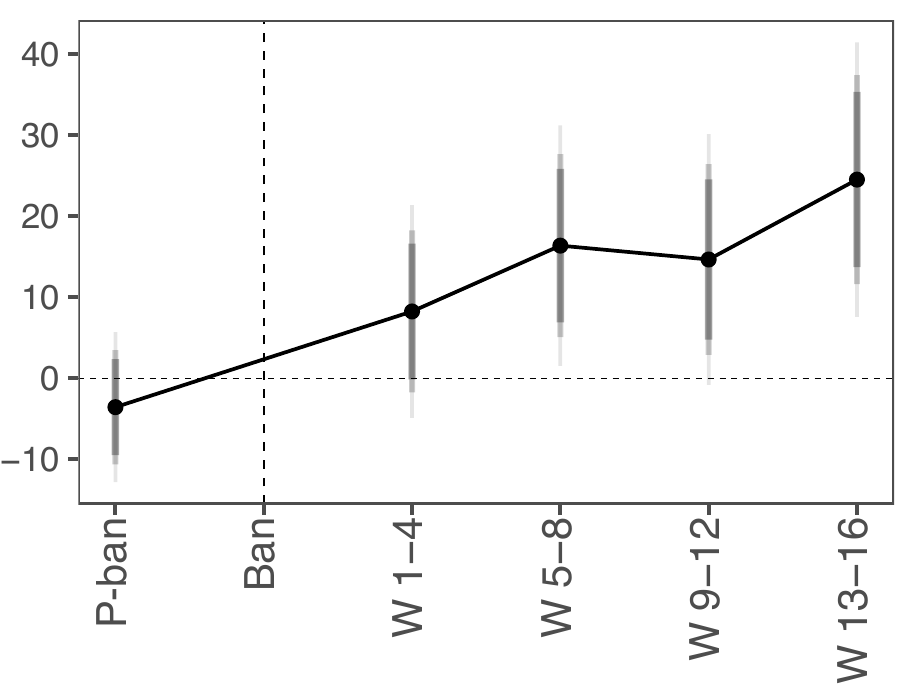}
    \end{subfigure}
    \caption{Estimated DiD effect of co-participation for toxicity (top row) and \engagement (bottom row) shown for r/The\_Donald (\cref{fig:did:thedonald}), r/GenderCritical (\cref{fig:did:gc}), and r/Incels (\cref{fig:did:incels}).
    To visualise the DiD effect, the plots are obtained via the causal model in \cref{eq:base_did} without fixed-effects.
    Effects are shown for the 5 four-weeks chunks (one for the pre-banning and four for the post-banning period).
    Error bars represent the 99\%, 95\% and 90\% CIs. Errors are clustered at user level. For details see \cref{table:coefficients} (top)}
    \label{fig:did}
\end{figure*}

We analyze the abovementioned differences with the DiD regression introduced in \cref{sec:methods}.
We quantify the effect of \emph{co-participation} in fringe platforms on \emph{Reddit} antisocial behavior.
To do so, we consider the four weeks before the ban as our pre-banning reference, and we group post-banning periods into four-weeks chunks.
We formalize this regression following \cref{eq:base_did}. 
Specifically, the dependent variables $Y_{it}$  are the toxicity ($T_{it}$) and the engagement ($E_{it}$) of each user $i$ at time $t$ grouped by $\Period_t$, i.e., four-weeks chunks.
The categorical variable $\Period_t$ refers to any of the five four-weeks chunks (one pre-banning and four post-banning). 

In \cref{fig:did}, we show the DiD effect, i.e., the difference in toxicity or \engagement between \ca and \ro net of pre-ban differences.
We observe that the pre-ban difference is zero, suggesting that the propensity score matching (see \cref{sec:psm}) adequately controls for pre-ban differences.
Most importantly, from \cref{fig:did}, we find that the DiD effects associated with each post-banning period increase over time for both toxicity and group engagement.
The four DiD coefficients (reported in \cref{table:coefficients}) increase with time, indicating that Co-Active users on Reddit become more toxic and engage more with controversial subreddits.
This result provides evidence that the adoption of antisocial behaviors by \ca users not only increases but also \emph{diverges} from that of \ro users.
However, we do not find evidence of such divergence in the cases of \engagement for r/GenderCritical (see \cref{fig:did:gc}(bottom)).
We speculate this might be because r/GenderCritical was banned with other $2,000$ subreddits. 
Such a mass ban might have caused most of the controversial communities associated with r/GenderCritical to get banned, too, thus limiting the ability of r/GenderCritical users to regroup.
For instance, the two subreddits r/TrueLesbians and r/Gender\_Critical, closely associated with r/GenderCritical, were jointly banned.

Interestingly, we notice that the DiD effect increases slowly for approximately eight weeks after the ban and starts to increase faster afterward. 
This finding is in line with the observation that users of banned subreddit may need time to become active (i.e., writing a post) on the fringe platform.
Indeed, 84\% of \ca users make their first post on the fringe platform between weeks 8 and 12 (vs. 10\% in weeks 1 to 7). 
Additionally, we note that only 6\% of users posted for the first time in weeks 13 to 18, suggesting that the analyzed timeframe is enough to capture the bulk of the user migration from Reddit to the fringe platform.

Additionally, we consider that co-participation may not necessarily be binary.
Instead, it can be considered as the volume of posts written by a user on the fringe platform. 
We formalize this \emph{activity} as the fraction of posts made on the fringe platform over all posts made by the user across platforms, i.e., Reddit and the fringe platform.
We hypothesize that increased activity in a fringe platform increases antisocial behavior on Reddit.
We then run a regression  where we substitute the binary variable $\Act_i$ with the user's activity.
Under this setting, we run this regression for r/The\_Donald to test if increased activity leads to an increase in antisocial behavior (i.e., toxicity and engagement). 
 We find that an increment of one percent of activity on the fringe platform translates into a toxicity increment of ~3.6 units and an engagement increment of ~5.4 units.

In synthesis, our results provide evidence that co-participation in fringe platforms affects users' antisocial behavior on Reddit.
We show that the toxic behavior of \ca users diverges over time from that of \ro users.  
In the following section, we estimate the rate of divergence.

\subsection{Divergence Analysis} 

We expand the regression of  \cref{eq:base_did} such that it considers the following dependent variables:
(i) $t$, an integer variable taking values in $[-18, +18]$;
(i) $\text{Period}_t$ a discrete variable indicating before and after ban periods; 
(ii) a fixed-effect $u_i$ for each user $i$.

We then model the dependent variable $Y_{it}$ as the log of $T_{it}$ and $E_{it}$.
This transformation addresses two issues observed in the data: the skewness of the dependent variable and a non-linear increment of antisocial behavior over time.
We formalize this regression as:
\begin{equation}\label{eq:temp_did}
\begin{split}
	\log(Y_{it}) =& \beta_0 + \beta_1 \Act_i+ \beta_2 \Period_t  + \beta_3 t +\\
   +& \beta_4 \Act_i \Period_t + \beta_5 \Act_i t +\\
   +&\beta_6 \Period_t t
       + \beta_7 \Act_i \Period_t t + u_i+ \varepsilon_{it}\,.
\end{split}
\end{equation}

\begin{figure*}[t!]
    \centering
    \begin{subfigure}{.33\linewidth}\centering
    \caption{r/The\_Donald}\label{fig:did_t:thedonald}
    \includegraphics[scale=.6]{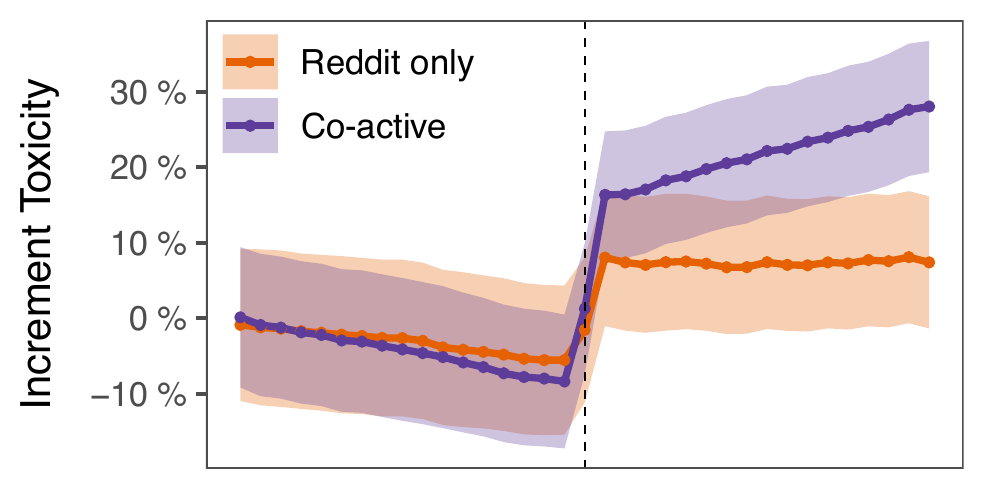}\\
    \includegraphics[scale=.6]{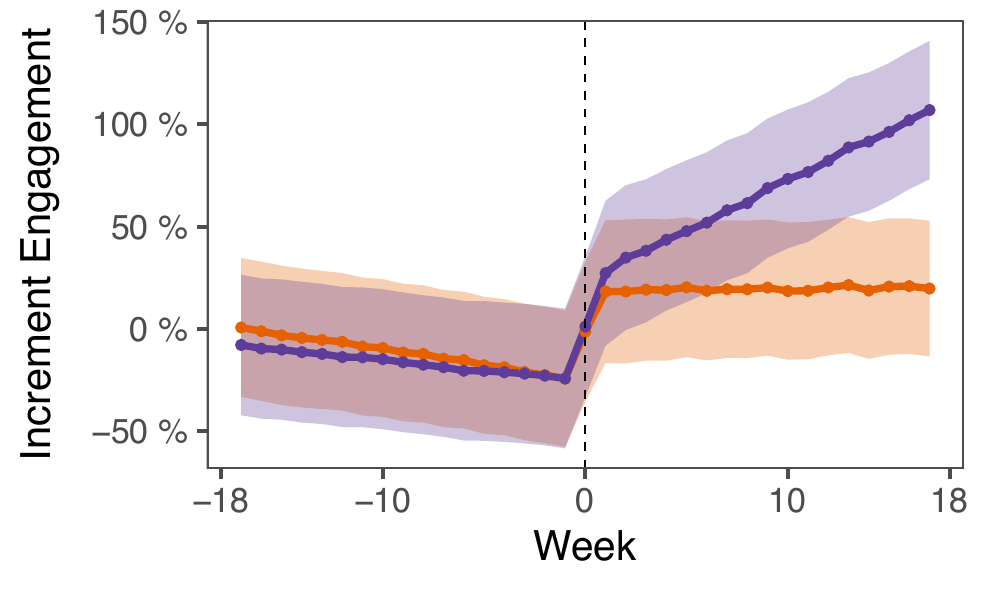}
    \end{subfigure}\hfill
   	\begin{subfigure}{.33\linewidth}\centering
   	\caption{r/GenderCritical}\label{fig:did_t:gc}
    \includegraphics[scale=.6]{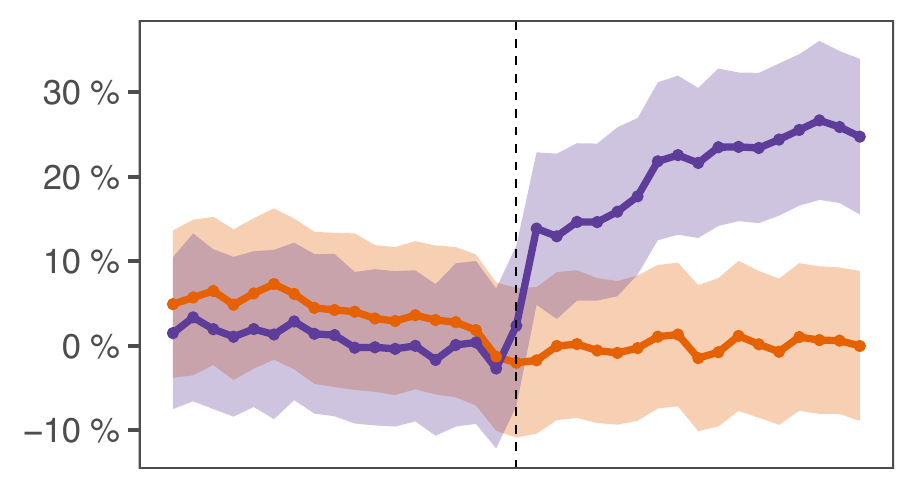}\\
    \includegraphics[scale=.6]{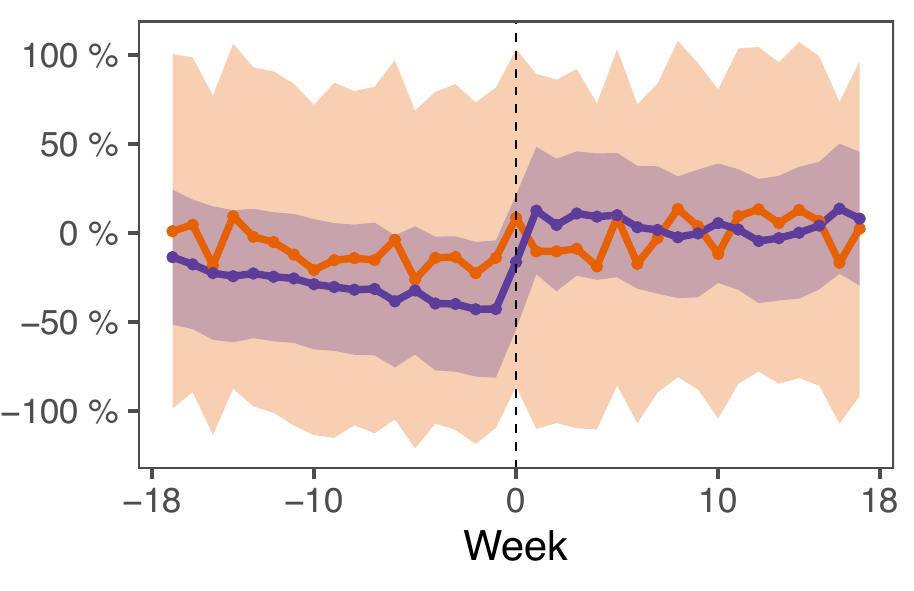}
    \end{subfigure}\hfill
	\begin{subfigure}{.33\linewidth}\centering
	\caption{r/Incels}\label{fig:did_t:incels}
    \includegraphics[scale=.6]{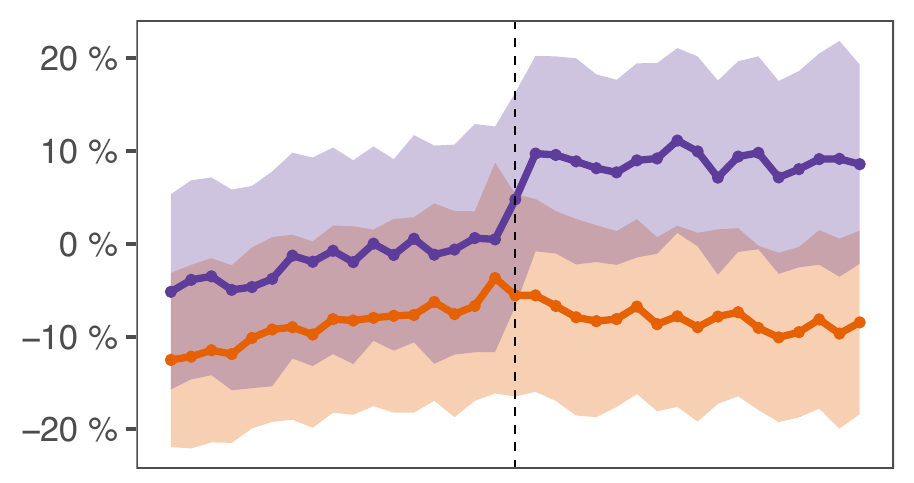}\\
    \includegraphics[scale=.6]{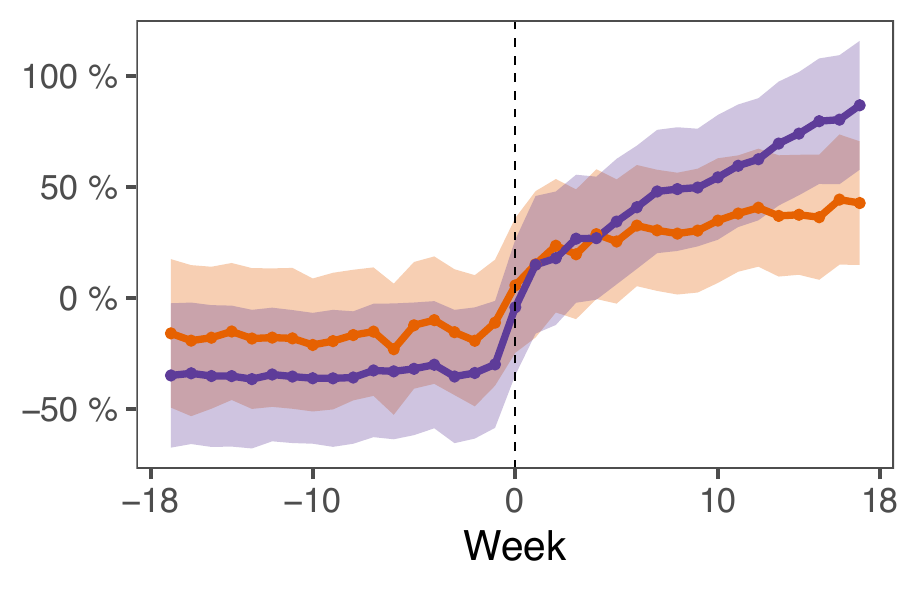}
    \end{subfigure}
    \caption{Divergence of toxicity (top row) and engagement (bottom row) for \ca (purple) and \ro users (orange).
    The average predicted relative increase for toxicity and \engagement are shown for r/The\_Donald (\cref{fig:did_t:thedonald}), r/GenderCritical (\cref{fig:did:gc}), and r/Incels (\cref{fig:did_t:incels}).
    The shaded areas represent the 95\% CIs. For further details, see \cref{table:coefficients} (bottom).}
    \label{fig:did_t}
\end{figure*}

The coefficient $\beta_7$ captures the weekly percentage increase in antisocial behaviors of \ca over \ro users.
Therefore, $\beta_7$ measures the divergence between the two groups.
The results are reported in \cref{table:coefficients}(bottom).
In \cref{fig:did_t}, we show the fitted models for the three subreddits. 
\Cref{fig:did_t:thedonald} top and bottom shows the model fitted on r/The\_Donald.
We observe that \ca users diverge consistently from \ro users in toxicity and engagement.
In particular, we find that the increase in toxicity and engagement for  \ca users exceeds that of \ro users by 2\% and 6\% \emph{per week}, respectively.
In r/Incels, the results for \engagement are qualitatively similar to those of r/The\_Donald.
In the case of r/GenderCritical, we find that the effect size on toxicity is similar to the one observed for r/The\_Donald, albeit less significant. 
We hypothesize that the lower statistical significance results from the smaller sample size of r/GenderCritical ($3,263$ samples against $50,628$).
We do not find evidence of an effect of co-participation on the toxicity of r/Incels users.
This last result is not surprising as users of r/Incels continued their activity on r/brainincels.
r/braincels allowed users of r/Incels to maintain their antisocial behavior, therefore mitigating the effect of the banning. 
Similarly, we do not find evidence of an effect of participation on the engagement of r/GenderCritical users.
Again, due to the mass ban of 2020, we argue that r/GenderCritical users could not find subreddits hosting similar groups.

With this analysis, we provide statistical evidence that the antisocial behavior of co-participating users not only sharply increases immediately after the ban but keeps growing at a higher rate than that of Reddit-Only users.
This differential growth results in a steady divergence in antisocial behavior once co-active users start participating in the highly toxic fringe platforms.

\section{Discussion}\label{sec:discussion}
Users on fringe platforms are exposed to a more toxic environment, which may spill over onto mainstream social media.
To test whether such spillover exists, we investigate if co-active users---active on both fringe platforms and mainstream social media---become more toxic on the mainstream platform after joining a fringe platform.
We study three controversial communities (r/The\_Donald, r/GenderCritical, and r/Incels) on Reddit by combining two well-established quasi-experimental methods: propensity score matching (PSM) and difference-in-differences regression (DiD).

We find that co-active users exhibit consistent and increased antisocial behavior on Reddit.
This increase diverges from users of the same banned community posting only on Reddit.
In particular, we find that the effect of co-participation intensifies with time and activity in fringe platforms. 
To support the causal interpretation of our results, we controlled for user-level characteristics by adding user-level fixed effects.

Our results shed light on the relations between fringe and mainstream social media.
While stakeholders of mainstream social media may consider the out-migration of users exhibiting antisocial behavior to be in their best interest, assuming that their platform and the fringe platform users migrated to are independent, our study reveals that co-active users act as a channel through which antisocial behavior on fringe platforms spills back onto mainstream social media.

While previous work has suggested that users ``adjust'' to toxicity levels of existing communities on Reddit~\cite{rajadesingan2020quick}, our results indicate that users exposed to toxic environments on fringe platforms will act similarly on the mainstream platform.

\paragraph{Implications.}
Our results have two critical implications for platform stakeholders.
First, they suggest that community-level bans are no silver bullets:
(i) community-level bans disproportionally increase antisocial behavior on fringe platforms~\cite{horta2021platform}, 
and (ii) this antisocial behavior spills over onto the mainstream platform, limiting the efficacy of such a moderation policy.
Stakeholders should thus be judicious with community-level banning.
Second, our results provide a clear target to reduce unintended within-platform consequences of community-level bans: \emph{co-active users}.
Platforms could develop more sophisticated interventions that remove problematic communities \emph{and} discourage co-activity.
For example, when banning communities like those studied, Reddit could \textit{also} apply sanctions to their users, such as reducing the visibility of their posts.
This friction could decrease co-activity levels and, as a consequence, mitigate the spillover of antisocial behavior.

\paragraph{Limitations.}
First, in our data, we may have incorrectly labeled as \ro users those \ca users that changed their usernames across platforms.
This results in a comparison of a group (\ca users) where every user co-participate on both platforms with a group (\ro users) where some users might have posted on the fringe platform under a different username.
Assuming that mislabelled \ca users do not behave in the opposite direction of correctly labeled \ca users, such a mislabelling can only decrease the observed effects.
This makes our results a lower bound of the true effect.

Second, our causal conclusions apply to \ca users who maintain their username when becoming active on the fringe platform. 
Nonetheless, we stress that the number of users that kept usernames across platforms was consistent.
For example, in thedonald.win, over 20\% of users could be matched to users in r/The\_Donald~\cite{horta2021platform}.
They even had a booking system in place to ensure that users could keep their usernames~\cite{donaldpost1}.
Further, users who kept their usernames are more active than the ones that did not~\cite{horta2021platform,zannettou2018origins}.
Therefore, studying these users is particularly important, as they  disproportionally impact our information ecosystem.

\paragraph{Future work.} We have investigated the effect of co-participation in fringe platforms on users' behavior on mainstream platforms.
Future work could investigate why users become active on fringe platforms after a ban , e.g., push and pull factors such as their position in the social network \cite{newell2016user, Russo2022UnderstandingOM}.
Also, our findings indicate that users who post more frequently on fringe platforms tend to exhibit more antisocial behavior on mainstream platforms. 
In contrast, future research may explore the effect on the behavior on mainstream platforms of users' exposure to fringe content (i.e., reading posts).

\section{Ethics Statement}

A positive outcome of our research is that it can help mainstream platforms design policies to mitigate the spillover of antisocial behavior.
For example, a platform might introduce automatic labeling of communities similar to banned ones, allowing users to make more informed decisions about their participation.
However, our findings may also be used to justify turning a blind eye to problematic communities, citing spillover concerns.
For example, a platform might tolerate abusive behavior in isolated communities rather than risk the spillover of that behavior to the wider platform following a ban.
We primarily use publicly available data that does not require user consent.
We collect data from the fringe platforms because it is an integral part of this research.
We do not use any personally identifiable information (PII) from the dataset, and we do not make any inferences about individual users.
Similarly, we do not name any other subreddits or users associated with the banned communities.
We confirm that we have read and abide by the AAAI code of conduct.
\begin{table*}[t!]
\small
\centering
\begin{tabular}{l D{.}{.}{2.6} D{.}{.}{2.6} D{.}{.}{2.6} D{.}{.}{2.8} D{.}{.}{2.6} D{.}{.}{2.6}}
\toprule
\multicolumn{7}{c}{DiD Analysis (with users fixed effects)}\\
\midrule
& \multicolumn{2}{c}{r/The\_Donald} & \multicolumn{2}{c}{r/GenderCritical} & \multicolumn{2}{c}{r/Incels}\\
\midrule
 & \mc{Toxicity} & \mc{Engagement} & \mc{Toxicity} & \mc{Engagement} & \mc{Toxicity} & \mc{Engagement} \\
\midrule
Coactive         & 0.625       & 92.301^{***} & -2.372       & -2.449   & -17.974      & 9.998        \\
                         & (7.177)     & (6.480)      & (4.273)      & (11.701) & (11.142)     & (19.095)     \\
Coactive:Period1 & 2.302^{***} & 3.668^{***}  & 7.748^{***}  & 2.039    & 8.270^{***}  & 4.292        \\
                          & (0.549)     & (0.764)      & (1.810)      & (3.339)  & (1.658)      & (3.826)      \\
Coactive:Period2 & 1.737^{**}  & 2.518^{**}   & 6.186^{***}  & 4.227    & 7.050^{***}  & 11.910^{**}  \\
                         & (0.552)     & (0.782)      & (1.748)      & (3.362)  & (1.677)      & (4.031)      \\
Coactive:Period3 & 3.046^{***} & 7.825^{***}  & 9.001^{***}  & -3.050   & 7.481^{***}  & 8.201^{*}    \\
                         & (0.564)     & (0.801)      & (1.763)      & (3.320)  & (1.698)      & (4.010)      \\
Coactive:Period4 & 6.644^{***} & 8.196^{***}  & 10.333^{***} & 3.186    & 7.303^{***}  & 18.267^{***} \\
                         & (0.572)     & (0.816)      & (1.782)      & (3.555)  & (1.634)      & (4.111)      \\
\midrule
\multicolumn{7}{l}{\scriptsize{\emph{Controls}}}\\
\midrule
(Intercept)              & 14.260^{*}  & -5.927       & 16.394^{***} & 2.909    & 34.919^{***} & 16.667       \\
                         & (6.417)     & (4.582)      & (3.218)      & (10.486) & (9.934)      & (16.684)     \\
Period & \mc{Yes} & \mc{Yes}& \mc{Yes}& \mc{Yes}& \mc{Yes}& \mc{Yes}\\
User Fixed Eff. & \mc{Yes} & \mc{Yes} & \mc{Yes} & \mc{Yes} & \mc{Yes} & \mc{Yes}\\
\midrule
R$^2$                    & 0.351       & 0.485        & 0.391        & 0.582    & 0.426        & 0.532        \\
Adj. R$^2$               & 0.291       & 0.448        & 0.341        & 0.365    & 0.359        & 0.474        \\
Num. obs.                & 23158       & 23158        & 1783         & 1783     & 3129         & 3129         \\
\end{tabular}
\begin{tabular}{l D{.}{.}{2.6} D{.}{.}{2.6} D{.}{.}{2.6} D{.}{.}{2.8} D{.}{.}{2.6} D{.}{.}{2.6}}
\midrule
\multicolumn{7}{c}{Divergenge Analysis (with users fixed effects)}\\
\midrule
Coactive               & -0.300       & 3.788^{***}   & 0.095         & -0.216        & -0.501         & 0.646       \\
                               & (0.193)      & (0.306)       & (0.172)       & (0.623)       & (0.306)        & (0.523)     \\
Banning                    & 0.302^{***}  & 0.565^{***}   & -0.031        & 0.195         & -0.064         & 0.494^{***} \\
                               & (0.018)      & (0.031)       & (0.058)       & (0.129)       & (0.050)        & (0.121)     \\
t                              & -0.007^{***} & -0.020^{***}  & -0.009^{*}    & -0.019^{*}    & 0.009^{*}      & 0.008       \\
                               & (0.001)      & (0.002)       & (0.004)       & (0.009)       & (0.004)        & (0.008)     \\
Coactive:Banning   & 0.328^{***}  & 0.077^{\cdot} & 0.437^{***}   & 0.302         & 0.392^{***}    & 0.123       \\
                               & (0.023)      & (0.043)       & (0.084)       & (0.199)       & (0.072)        & (0.161)     \\
Coactive:t             & -0.005^{***} & 0.006^{*}     & 0.006         & 0.004         & -0.007         & 0.005       \\
                               & (0.001)      & (0.003)       & (0.006)       & (0.013)       & (0.005)        & (0.011)     \\
Banning:t                  & 0.009^{***}  & 0.021^{***}   & 0.011^{\cdot} & 0.022^{\cdot} & -0.009^{\cdot} & 0.023^{*}   \\
                               & (0.002)      & (0.003)       & (0.006)       & (0.013)       & (0.005)        & (0.012)     \\
Coactive:Banning:t & 0.022^{***}  & 0.060^{***}   & 0.012         & -0.021        & 0.004          & 0.061^{***} \\
                               & (0.002)      & (0.004)       & (0.009)       & (0.020)       & (0.008)        & (0.015)     \\
\midrule
\multicolumn{7}{l}{\scriptsize{\emph{Controls}}}\\
\midrule
(Intercept)                    & 2.859^{***}  & -0.096        & 2.408^{***}   & 1.123^{*}     & 2.922^{***}    & 0.938^{*}   \\
                               & (0.157)      & (0.215)       & (0.122)       & (0.569)       & (0.239)        & (0.381)     \\
User Fixed Eff. & \mc{Yes} & \mc{Yes} & \mc{Yes} & \mc{Yes} & \mc{Yes} & \mc{Yes}\\
\midrule
R$^2$                          & 0.373        & 0.406         & 0.358         & 0.476         & 0.367          & 0.442       \\
Adj. R$^2$                     & 0.346        & 0.384         & 0.331         & 0.336         & 0.325          & 0.407       \\
Num. obs.                      & 50628        & 50628         & 3263          & 3263          & 5433           & 5432        \\
\bottomrule
\multicolumn{7}{l}{\scriptsize{$^{***}p<0.001$; $^{**}p<0.01$; $^{*}p<0.05$; $^{\cdot}p<0.1$}}
\end{tabular}
\caption{Summary of results. (top) Coefficient estimates with fixed effects for the DiD analysis. (bottom) Coefficient estimates and standard errors for the divergence analysis. In the second regression, the response variables have been log-transformed.}
\label{table:coefficients}
\end{table*}

\appendix

\begin{figure*}[t]
    \centering
    \begin{subfigure}{.33\linewidth}\centering
    \caption{r/The\_Donald}\label{fig:data:thedonaldlove}
    \includegraphics[scale=.47]{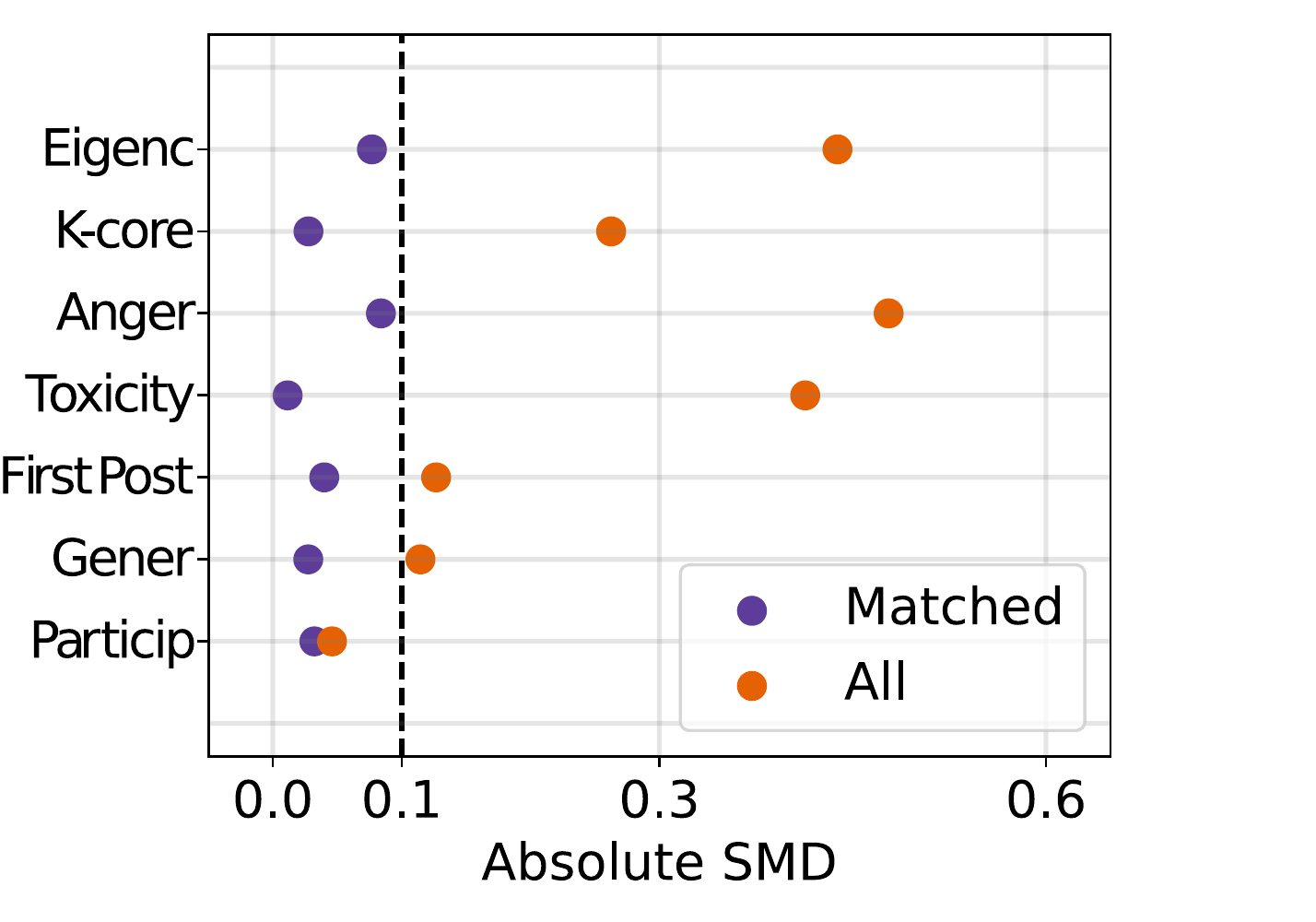}
    \end{subfigure}\hfill
   	\begin{subfigure}{.33\linewidth}\centering
   	\caption{r/GenderCritical}\label{fig:data:gclove}
    \includegraphics[scale=.47]{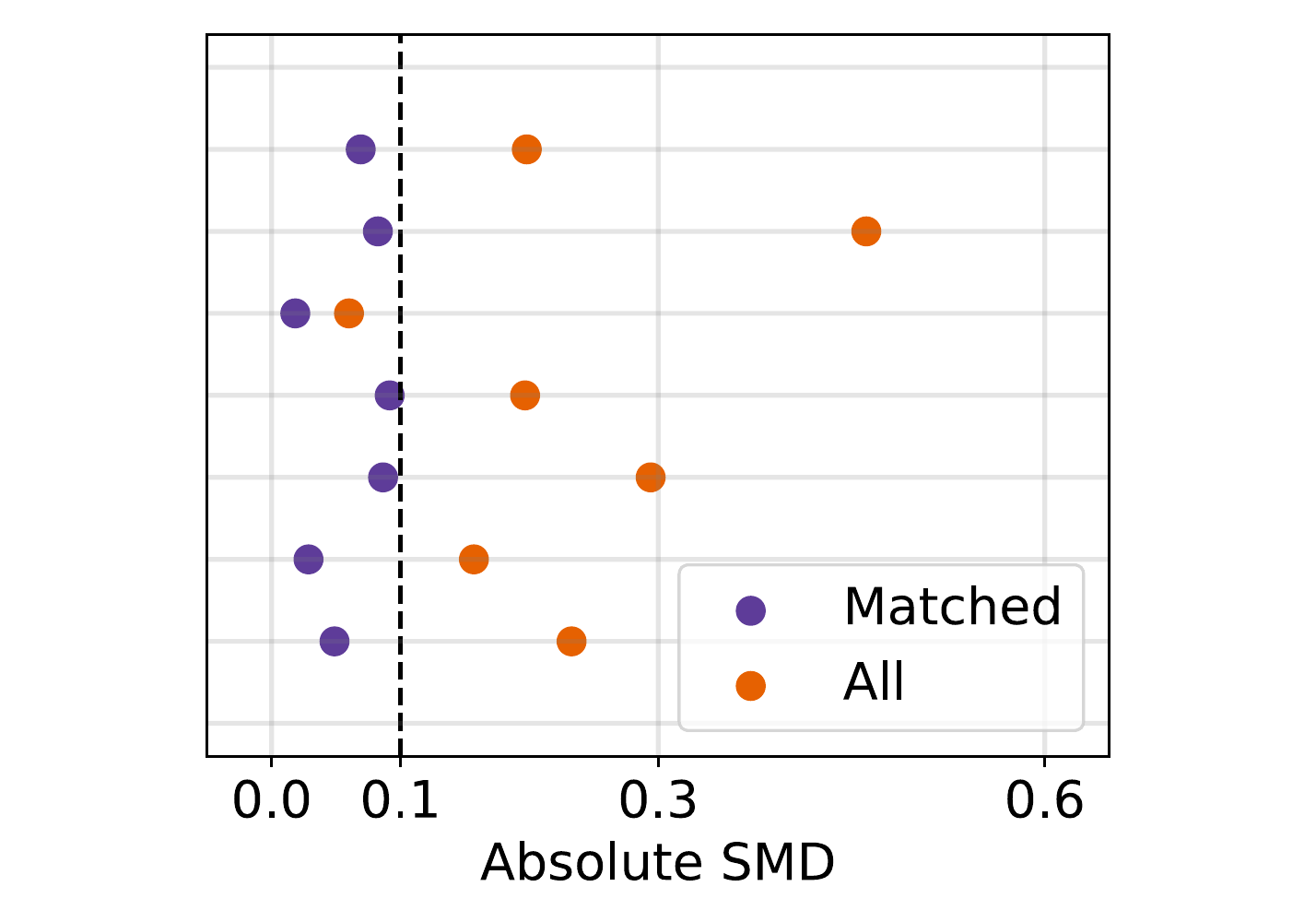}
    \end{subfigure}\hfill
	\begin{subfigure}{.33\linewidth}\centering
	\caption{r/Incels}\label{fig:data:incelslove}
    \includegraphics[scale=.47]{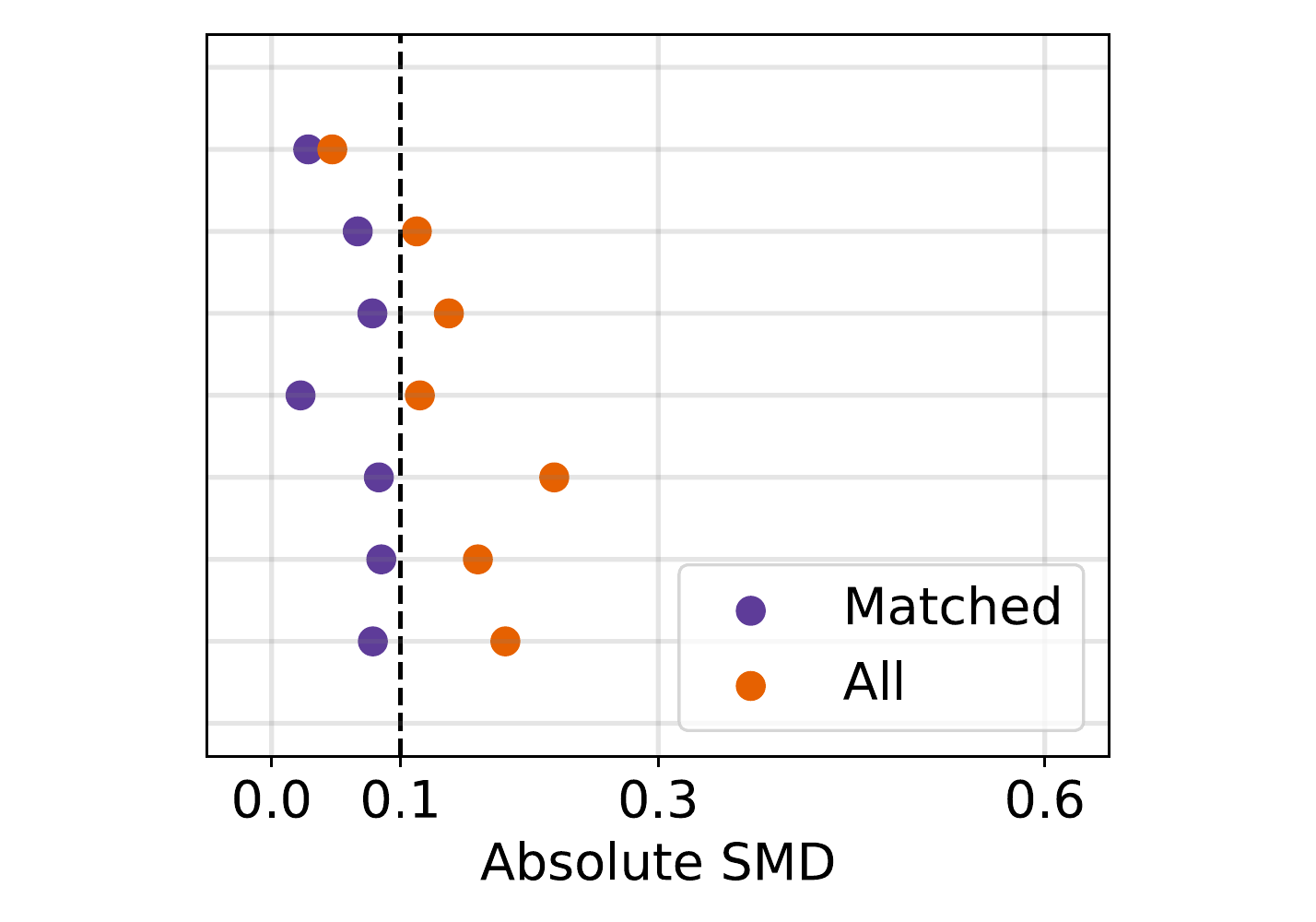}
    \end{subfigure}
    \caption{
    Absolute standardized mean differences (SMD) between \ca and \ro before and after matching for all the covariates used for the propensity score matching.
    The SMDs have been computed for all three considered subreddits  r/The\_Donald (\cref{fig:data:thedonaldlove}), r/GenderCritical (\cref{fig:data:gclove}), and r/Incels (\cref{fig:data:incelslove}).}
    \label{fig:asmd}
\end{figure*}

\section{Methodological Details}\label{sec:appendix}

\paragraph{Subreddits similarity scale.}
To create a similarity scale between subreddits, we map the similarity score to $[-1,+1]$, where 1 represents the highest similarity to the considered community.
To do so, we follow the method proposed by \citet{Waller2020QuantifyingSO}.
We consider our focal subreddits r/The\_Donald, r/GenderCritical, and r/Incels and their opposites r/HillaryClinton, r/asktransgender, and r/feminists, respectively.
The opposite subreddits were chosen by identifying those that were similar in all aspects except for one specific characteristic, different from the focal subreddits \subreddits. For example, r/HillaryClinton is considered the opposite of r/The\_Donald because, while both host discussions about politics, r/HillaryClinton is on the opposite side of the political spectrum compared to r/The\_Donald.
Given a subreddit, we define as ``relevant'' all other subreddits where at least ten users of the subreddit posted at least five times.
We then define a graph for each focal subreddit, where the nodes consist in a subreddit; either the focal subreddit (e.g., r/Incels), its opposite (e.g., feminists), and all relevant subreddits for the focal subreddit and its opposites.
We draw a weighted edge between two nodes if the corresponding subreddits share at least five users, with the weight equal to the number of users shared.
We train the Node2Vec~\cite{Grover2016node2vecSF} algorithm on each graph to get embeddings of each subreddits of the graphs.
Finally, we use the cosine similarity to obtain the similarity between our considered subreddits and those included in each graph.
Using this similarity scale, we compile a list of the top \emph{k} most similar subreddits to r/The\_Donald, r/GenderCritical, and r/Incels.

\paragraph{Validation of Similarity Scale.}
To validate the subreddit similarity scale, we refer to the concept of convergent validity.
This concept measures the correlation between our similarity scale and other measures based on the same construct.
We use the only publicly available subreddit embeddings by \citet{Waller2020QuantifyingSO} for this comparison.
The embeddings from \citet{Waller2020QuantifyingSO} are not explicitly trained toward finding similarities between specific communities.
Nevertheless, they provide a general measure of subreddit similarity.
We calculate Spearman's rank-order correlation between the 1000 subreddits most similar to r/The\_Donald, r/GenderCritical, and r/Incels  according to our and \citet{Waller2020QuantifyingSO} ranking.
We find a significant (p $<$ 0.05) moderate correlation (0.64) between the two.
This result corroborates that our similarity scale successfully measures similarity to r/The\_Donald, r/GenderCritical, and r/Incels.

\paragraph{Manual Annotation}
Finally, to compute the engagement in controversial groups for a user $i$, we need to individuate the top-K most similar communities to the subreddit associated with user $i$, i.e., \subreddits.
Three authors annotated the top 100 most similar subreddits to determine which ones hosted discussions similar to one of the focal subreddits (\subreddits). 
They labeled each subreddit as ``similar'' if its discussion was similar to the focal subreddit's and ``not similar'' otherwise. 
We measured the inter-annotator agreement, which resulted in a score of 0.82.
The annotators found that 97\%, 96\%, and 99\% of the subreddits labeled as ``similar'' to r/The\_Donald, r/GenderCritical, and r/Incels, respectively, were in the first \emph{50 most similar subreddits} according to the similarity scale described above.
This lead us to choose $k=50$ in our analyses.

\section{Propensity Score Matching}\label{app:propensity_score}

\paragraph{PSM Covariates.}
We define the covariates used to perform the PSM to match \ca and \ro users.
\begin{itemize}
    \item \textbf{Participation}: We compute \emph{participation} following the approach of \citet{phadke2022pathways}.
We define the participation of a user $i$ at time $t$ as $p_{it}=\frac{n_{s_j}\text{sim}(s_b,s_j)}{N_i}$.
Where $n_{s_j}$ is the number of comments made on the  subreddit $s_j$, $\text{sim}(s_b,s_j)$ is the similarity between the embeddings of the banned subreddit $s_b$, (e.g., r/Incels) and $s_j$ computed as described above.
$N_i$ is the total number of comments on Reddit of user $i$. 
$p_{it}$ is bounded between 0 to 1, with higher scores indicating high participation in the banned subreddits discussion.
\item \textbf{Generality Score:} The generality score is a measures defined by \citet{waller2019generalists}. It is bounded between -1 and +1. Users with a score of +1 post in multiple and diverse subreddits. Users that have a score of -1 are instead specialists. The generality score is the average cosine similarity between the embeddings of subreddits in which a user $i$ is active and his center of mass, weighted by
the number of contributions by the community. $ i$'s center of mass is defined as the weighted average of the embeddings of the subreddits in which $i$ participated.
\item \textbf{First Day post}: The difference in days between the date of the first post and the banning date of the subreddit. 
\item \textbf{Toxicity}: We compute the weekly average toxicity of a user as described in \cref{sec:methods}
\item \textbf{Anger and Anxiety}: A count of anger and anxiety-related words identified via LIWC~\cite{tausczik2010psychological}.
\item \textbf{K-Core centrality}: We build a communication network using only the banned subreddits. Nodes are users, and edges exist if a user has answered another user's post more than five times. The k-core centrality is the subgraph of nodes in the k-core but not in the (k+1)-core.
\item \textbf{Eigencentrality}: Using the same network we used to compute the k-core centrality, we compute the eigencentrality of each node. 
\end{itemize}

\paragraph{Robustness of PSM}
We have evaluated the robustness of our results against different matching algorithms.
Specifically, we have tested: nearest neighbor, genetic matching, and coarsened exact matching (CEM). 
We found that our results were independent of the choice of the matching algorithm and decided to use the nearest neighbor algorithm, arguably the simplest and the default in the widely used \emph{MatchIt} package.%
\footnote{\url{https://cran.r-project.org/web/packages/MatchIt/index.html}}
To show the quality of our matching, we show in \cref{fig:asmd} `love plots' for the absolute standardized mean differences of all covariates used in the propensity score matching.
Specifically, we show the values of the absolute standardized mean errors (ASMD) before and after matching for all three subreddits we analyzed. 
The ASMD is below the standard 0.1 threshold for all covariates.

\end{document}